\documentstyle[12pt,preprint,tighten,floats,aps,epsf,psfig]{revtex} 
\def\simlt{\stackrel{<}{{}_\sim}} 
\def\simgt{\stackrel{>}{{}_\sim}} 
\tighten 
\preprint{IC/99/13}
\begin{document} 
\draft \title{Distinguishing the Higgs scalars of NMSSM and $\mbox{Z}^{\prime}$ models 
for large Higgs trilinear couplings} 
\author{D. A. Demir} 
\address{The Abdus Salam International Centre
for Theoretical Physics, I-34100 Trieste, Italy} 
\date{\today} 
\maketitle 
\begin{abstract} 
Two well-known extended supersymmetric models, $Z^{\prime}$ models and NMSSM, 
are comparatively analyzed in the limit of large trilinear Higgs couplings.
The two models are found to have distinguishable Higgs spectra at both tree- and 
loop- levels. Higgs production through Bjorken processes at an $e^{+}e^{-}$ collider
is shown to discriminate between the two models.\\ 
\end{abstract} 
\newpage 
\section{Introduction} A proper understanding of the basic mechanism of the
electroweak symmetry breaking is one of the central problems in particle 
physics. If the fundamental particles are to remain weakly
interacting up to high energies the symmetry breaking sector of 
the model should contain one or more scalar Higgs bosons with masses
of the order of weak scale $\sim G_{F}^{-1/2}$. The simplest 
mechanism for the breaking of the electroweak symmetry is realized in
the SM where one scalar field remains in the spectrum, 
manifesting itself as the physical Higgs particle $H$. Though the scalar
sector of the SM is simple enough to predict just one 
Higgs boson, it has been criticized from various theoretical aspects, the most
common of which is the quadratic divergence of the Higgs mass. 
This naturalness problem has triggered the study of supersymmetric
models.  MSSM, being the simplest of such models, lacks an 
explanation for the origin and scale of the bilinear Higgs coupling mass,
which is commonly known as the $\mu$ problem. In connection 
with this last point, extended supersymmetric models, in which bilinear
Higgs coupling mass $\mu$ is related to the vacuum expectation 
value of some SM gauge singlet, have been proposed. In these extended
models, one extends either merely the Higgs sector (NMSSM) \cite{nmssm} , 
or both Higgs sector and the SM gauge group ($Z^{\prime}$
models) \cite{lang,lang2,zprime}, and these generally lead to a larger 
Higgs spectrum.Thus, in accelerator searches of the Higgs, even if
detected, it will still be a challenging issue to determine what 
kind of model it implies. Consequently, it is desirable to
investigate the possibility of discriminating between these models 
given that some spectrum of Higgs particles are observed at some
($e^{+}e^{-}$) collider.

High precision electroweak data indicate a preference for Higgs boson to 
have a mass within a factor of 2 or so of 140 GeV \cite{preci}. However, 
weak (logarithmic) dependence of the theoretical predictions on the Higgs 
boson mass as well as the intrinsic uncertainities in the electroweak 
observables prevent to derive stringent predictions for the Higgs mass, 
and literally almost the entire SM Higgs boson range (up to and above $\sim$ 1 TeV) 
must be swept for the Higgs search \cite{sm1}. From the direct searches, 
four LEP experiments put the lower bound of 95 GeV for the Higgs mass 
\cite{LEP}. The dominant production mechanism for Higgs boson within the 
LEP2 reach is the Higgs strahlung process, $e^{+}e^{-}\rightarrow Z H$, 
in which Higgs boson is emitted from a virtual $Z$ line. This process 
has already been analyzed in SM \cite{sm} and MSSM \cite{mssm}, 
including the radiative corrections. One notes that for 
$\nu_{e}\bar{\nu}_{e}$ 
and $e^{+}e^{-}$ in the final state, fusion process where Higgs boson is 
formed in $WW$ and $ZZ$ t-channel collisions will interfere with the 
Higgs- strahlung amplitude. 

In this work we analyze NMSSM and $Z^{\prime}$ models comparatively
with particular emphasis on their CP--even Higgs spectra. To have 
direct effects of these particles at the weak scale one demands 
the gauge singlet scalar to have a vacuum expectation value (VEV) around
the weak scale. Such a breaking scheme leading relatively light
particles (including the $Z^{\prime}$ boson itself) has already
been shown to exist in $Z^{\prime}$ models \cite{lang}. This occurs 
when the trilinear coupling of the Higgs doublets and the singlet 
in the soft supersymmetry breaking terms becomes large compared to
the soft masses, and leads to appoximately equal VEV's for both 
doublets and the singlet. This type of vacuum state satisfies 
the existing constraints on the mixing between the $Z^{0}$ and 
$Z^{\prime}$, and produces a spectrum of scalars at the weak scale.

However, in NMSSM such Higgs trilinear coupling driven minima, though 
can exist, are not required by some phenomenological requirement \cite{nmssm}
as in the $Z^{\prime}$ models \cite{lang}. In $Z^{\prime}$ models for singlet 
to have large VEV one needs the singlet mass-squared to be large negative. In NMSSM,
in addition to this mechanism, one can allow for the singlet trilinear coupling 
to be large negative to have a large singlet VEV. The opposite limit, namely,
the requirement of  having both singlet and the doublet VEV's to be around the weak 
scale could be satisfied by choosing the mass-squareds of the Higgs fields 
appropriately, in particular, letting them be sufficiently larger than both 
singlet cubic coupling and Higgs trilinear coupling, in absolute magnitude. 
However,this requirement can also be met by choosing Higgs trilinear coupling 
large enough compared to soft masses and singlet cubic coupling. 

In this work we will perform a comparative study of NMSSM and $Z^{\prime}$ 
models in the same kind of minimum induced by the relatively large values of the 
Higgs trilinear coupling. In particular, we dicuss the CP--even spectra, and 
investigate their collider signature, both at tree-- and loop-- levels, by
analyzing the Higgs production through the Bjorken processes. 

In Sec.2 we derive the couplings and  masses of the scalars, and 
specify all of the  relevant properties of NMSSM and $Z^{\prime}$ 
models. We also discuss vacuum stability and scalar spectrum against
radiative corrections.

In Sec.3 we give a comparative discussion of the electron-positron 
annihilation to four-fermion final states through Higgs-strahlung in both 
NMSSM and $Z^{\prime}$ models with reference to the corresponding SM 
expressions. 

In Sec. 4 we conclude the work.
\section{Scalar sectors of NMSSM and  $Z^{\prime}$ models}
We consider a supersymmetric model whose scalar sector consists of two 
Higgs doublets $\hat{H}_{1}$, $\hat{H}_{2}$, and a SM-singlet $\hat{S}$.
NMSSM is such a supersymmetric model whose scalar sector is spanned by 
these fields, and gauge symmetry is exactly that of the MSSM. In fact, it 
is specified by the superpotential
\begin{eqnarray}
W \ni h_{s}\,\hat{S}\,\hat{H}_{1}\cdot \hat{H}_{2} 
+\frac{1}{3}k_{s}\hat{S}^{3} +h_{t}\,\hat{U}^{c}\,\hat{Q}\cdot \hat{H}_{2} 
\end{eqnarray} 
where contributions of all fermions but the top quark are neglected as 
they are much lighter. Here $\hat{Q}$ and $\hat{U}^{c}$ are left-handed 
SU(2) doublet and singlet quark superfields. The cubic term in (1) is 
necessary to avoid the unwanted Peccei-Quinn symmetry. However, 
superpotential has still a $Z_{3}$ symmetry which, when spontaneously 
broken, causes serious problems about domain walls \cite{wall}. This 
problem can be avoided by the addition of non-renormalizable terms of 
the form $(\hat{S}/M_{P})^{n}\hat{S}\hat{H}_{1}\cdot \hat{H}_{2}$ if gravity 
violates the $Z_{3}$ symmetry \cite{non-re}. Despite these problems, the 
superpotential (1) is sufficient to understand the basic implications of 
NMSSM for particle physics applications. 

Besides NMSSM, $Z^{\prime}$ models, which are those models having a low-energy supersymmetric 
extra U(1) factor, have also an extended scalar sector compared to MSSM. 
Unlike NMSSM, these models are devoid of such cosmological problems, and are 
predicted in string compactifications and $E(6)$ GUT's 
\cite{lang,zprime,dur}. In $Z^{\prime}$ models, not only the scalar sector but 
also the gauge sector is extended by an extra $U(1)$ factor with the 
coupling $g_{Y'}$. Consequently, all fields are charged under this 
group, and we make the charge assignment $Q_{1}$, $Q_{2}$, $Q_{S}$, $Q_{Q}$ 
and  $Q_{U}$ for $\hat{H}_{1}$, $\hat{H}_{2}$, $\hat{S}$, $\hat{Q}$, 
and $\hat{U}^{c}$, respectively. In this case, the superpotential is given by 
\begin{eqnarray}
W \ni h_{s}\,\hat{S}\,\hat{H}_{1}\cdot \hat{H}_{2} 
+h_{t}\,\hat{U^{c}}\,\hat{Q}\cdot \hat{H}_{2} 
\end{eqnarray}
where a cubic term is forbidden due to the extra Abelian group factor. Formally, 
one can obtain this superpotential by setting $k_{s}=0$ in (1). Although 
the superpotentials (1) and (2) formally differ only by the cubic term, 
resulting scalar potentials reveal explicitly the difference between the 
two models, through F- and D- terms. In fact, the most general 
representation for the scalar potential is given by  
\begin{eqnarray}
V&=&{m_{1}}^{2}\mid H_{1}\mid ^{2}
+{m_{2}}^{2}\mid H_{2}\mid ^{2} + {m_{S}}^{2}\mid S\mid
^{2}+ \lambda_{1}\mid H_{1}\mid ^{4} +\lambda_{2}\mid  
H_{2}\mid ^{4}\nonumber\\&+&\lambda_{S}\mid S\mid ^{4}+
\lambda_{12} \mid H_{1}\mid ^{2} \mid H_{2}\mid ^{2} +
\lambda_{1S} \mid H_{1}\mid ^{2} \mid S\mid ^{2} +
\lambda_{2S} \mid H_{2}\mid ^{2} \mid S\mid^{2}\nonumber\\&+&
\tilde{\lambda}_{12}\mid H_{1}^{\dagger} H_{2}\mid ^{2}+\lambda_{S12}(S^2 
H_{1}.H_{2}+H.c.)-h_{s}A_{s}(S H_{1}.H_{2} + H. 
c.)\nonumber\\&-&k_{s}A_{k}(S^{3} + H.c.) 
\end{eqnarray}
where the dimensionless $\lambda$ coefficients are listed in Table I for 
both models. In writing this potential we suppressed the contribution of 
the squarks for simplicity, however, when discussing the loop effects we 
will explicitly take them into account.

For later use we parametrise the vacuum expectation values of the 
Higgs fields as follows:
\begin{eqnarray}
<H_{1}>=\frac{1}{\sqrt{2}}\left(\begin{array}{c c}
v_{1}\\0\end{array}\right)\; , \;
<H_{2}>=\frac{1}{\sqrt{2}}\left(\begin{array}{c c}
0\\v_{2}\end{array}\right)\; , \;
<S>=\frac{v_{s}}{\sqrt{2}}
\end{eqnarray}
with real $v_{1}, v_{2}$ and $v_{s}$. All physical quantities of interest 
can be expressed in terms of parameters of the potential and these VEV's.
\subsection{Couplings of Vector Bosons to Fermions}
Before specializing to a particular minimum of the potential we derive
vector boson-fermion couplings as they are essentially independent of the 
scalar sector of the model. Since the vector boson sector of the 
$Z^{\prime}$ model is larger we analyze it in detail first, and then 
infer the necessary formulae for NMSSM. There are two massive 
neutral vector bosons in  $Z^{\prime}$ models, the usual $Z$ of the SM gauge 
group, and $Z^{\prime}$ of the extra $U(1)$ group, which mix through the 
mass-squared matrix:
\begin{eqnarray}
({\cal{M}}^{2})_{Z-Z'}=\left (\begin{array}{c c}
M_{Z}^{2}&\Delta^{2}\\\Delta^{2}&M_{Z'}^{2}\end{array}\right),
\end{eqnarray}
whose entries are given by
\begin{eqnarray}
M_{Z}^{2}&=&\frac{1}{4}G^2(v_{1}^2+v_{2}^2),\\
M_{Z'}^{2}&=&{g}_{Y'}^{2}(v_{1}^{2}Q_{1}^{2}+v_{2}^{2}Q_{2}^{2}
+v_{s}^{2}Q_{S}^{2}),\\
\Delta^{2}&=&\frac{1}{2}g_{Y'}\,G(v_{1}^2Q_{1}-v_{2}^2Q_{2}),
\end{eqnarray}
where $G^{2}=g_{2}^{2}+g_{Y}^{2}$. $Z-Z^{\prime}$ mixing angle, which 
is one of the most important parameters in $Z^{\prime}$ models, is 
defined by 
\begin{eqnarray}
\tan 2\theta=-\frac{ 2\Delta^{2}}{M_{Z'}^{2}-M_{Z}^{2}}
\end{eqnarray}
Diagonalization of $({\cal{M}}^{2})_{Z-Z'}$ leads to mass-eigenstates 
$Z_1$ and $Z_2$ with masses $M_{Z_{1}}$ and $M_{Z_{2}}$, respectively. 
The  $Z-Z^{\prime}$ mixing angle forms the mere sign of $Z^{\prime}$ models  
in LEP Z-pole data and, in fact, in this way it is constrained to be $\simlt 
10^{-3}$ \cite{lang}. Actually the way $\theta$ enters the LEP Z-pole 
observables can be seen through the $Z_{1}f\bar{f}$ couplings. For a 
fermion $f$, we define 
\begin{eqnarray}
\epsilon^{f}=I_{3}^{f}-Q_{em}^{f}\sin^{2}\theta_{W}\;\; , \; 
{{\epsilon}^{\prime}}^{f}=Q_{f}
\end{eqnarray}
where $I_{3}^{f}$ is the third component of the weak isospin, and 
$Q_{em}^{f}$ is the electric charge. Then coupling of $Z_{i}$ to a fermion 
line is given by the lagrangean 
\begin{eqnarray}
{\cal{L}}_{i}=\frac{G}{2}Z_{i}^{\mu}\bar{f}\gamma_{\mu}(g_{V}^{(i)}- 
\gamma_{5}g_{A}^{(i)})f
\end{eqnarray}
where vector coupling $g_{V}^{(i)}$ and  axial coupling $g_{A}^{(i)}$
are listed in Table II for $Z^{\prime}$ models. One notices that 
$g_{V}^{(i)}$ depends on the sum of the extra U(1) charges of the left--handed 
fermion doublet and right--handed SU(2) singlet, that is,  
$Q_{f_{L}}+Q_{f_{R}}$. Due to the gauge invariance of the superpotential 
this sum equals minus the extra U(1) charge of the associated Higgs 
doublet. Hence it is not possible to implement the leptophobic 
$Z^{\prime}$ models though it is required by the high energy precision 
data. Nevertheless, this problem can be sidestepped if lepton charges under 
extra U(1) group vanish, and there is no trilinear mass term in the 
superpotential for leptons, that is, they acquire masses through the 
non-renormalizable interactions as all light quarks are supposed to do 
\cite{faraggi}. Thus we rectrict our attention particularly to the top 
Yukawa coupling given in the superpotential (2). Unlike the vector 
couplings, axial couplings depend on the difference between the extra 
U(1) charges of the doublet and SU(2)-singlet fermions. Needless to say, 
in the limit of small mixing angles, $Z_{1}f\bar{f}$ couplings approach 
the corresponding SM ones. In NMSSM, one has only the standard 
$Z$ boson whose couplings can be obtained by setting $g_{V,A}^{(2)}=0 
,\;g_{Y'}=0$, and $\theta=0$ in Table II. NMSSM couplings can be compared 
with the results of \cite{nmssm}.

\subsection{Couplings of Vector Bosons to CP-even Higgs Bosons}
After obtaining the couplings of vector bosons to fermions, now we turn 
to the discussion of CP-even Higgs-vector boson couplings which are of 
fundamental importance in discussing the Higgs production through the 
Higgs-strahlung type processes. Evaluation of the couplings of Higgs bosons 
to vector bosons requires the minimization of the potential (3) after 
which one obtains the physical particle spectrum together with the 
necessary diagonalizing matrices. Each of the models under concern has 
its own phenomenological constraints to be satisfied. As discussed in the 
Introduction, in the case of NMSSM, one should prevent the creation of a
pseudoscalar Goldstone mode, which can easily be satisfied for  
non-vanishing $k_{s}A_{k}$. In the case of $Z^{\prime}$ models, 
however, one has  stringent constraints on the possible vacuum state, 
that is, the  $Z-Z^{\prime}$ mixing angle (9) should be $\simlt 10^{-3}$ 
\cite{lang}, and $Z^{\prime}$ mass must be $\simgt$ 600 GeV, as 
required by the recent  Tevatron direct search with leptonic final states 
\cite{tevatron}. As long as one considers the case of leptophobic 
$Z^{\prime}$ in accord with  the LEP Z-pole data, this latter condition 
can be relaxed. However, the former one is a non-trivial condition on 
the possible vacua of $Z^{\prime}$  models. In the next section we shall 
analyze the relevant vacuum state of the two models in detail.

For the models under concern, to have spectacular effects at LEP2 
and NLC energies, 
the breaking scale of the supersymmetry is expected to be around the weak scale. 
Indeed, when the supersymmetry is broken above $\sim$ TeV, at the weak 
scale one ends up with an effective  2HDM \cite{Ben,Ma} which carries only 
indirect information about the underlying model. Thus, in what 
follows we assume supersymmetry to be broken around the weak scale for both 
models under concern. This observation requires all three VEV's to be around 
the weak scale. In particular, $-m_{S}^{2}>>|m_{1,2}^{2}|$ is prohibited 
in both models as otherwise SM-singlet $S$ may acquire a large VEV. In 
NMSSM, in addition to this, one has to prevent $|k_{s}A_{k}| >> 
|h_{s}A_{s}|$ as the former one can induce a large VEV for $S$ through 
the cubic soft term. 

\subsubsection{Relevant minimum in $Z^{\prime}$ models}
We first discuss $Z^{\prime}$ models, and following it we turn to the 
discussion of NMSSM. In analyzing the $Z^{\prime}$ models the basic 
quantity of interest is the $Z-Z'$ mixing angle (9) which has to be small 
to satisfy the present phenomenological bounds \cite{lang}. An 
observation on (9) reveals that $Z-Z'$ mixing angle  $\theta$ can be made 
small either by choosing $M_{Z^{\prime}}> > M_{Z}$, or by forcing 
$\Delta^{2}$ itself to be small without constraining $M_{Z^{\prime}}$. 
While the former one requires SM-singlet VEV to be much larger than the 
doublet VEV's, the latter allows all three VEV's to be of the same order of 
magnitude, as required by the discussions at the end of the last 
subsection. One can realize small $\Delta^{2}$ when the charges 
of the Higgs doublets under extra U(1) are equal, and Higgs trilinear 
coupling $h_{s}A_{s}$ is larger than the other mass parameters  
\cite{lang,zprime}. We name that minimum of the potential (3) for which  
$h_{s}A_{s}$ is larger than the other mass parameters as Higgs trilinear 
coupling driven minimum (HTCDM) from now on. In fact, the potential possesses 
a HTCDM when $A_{s}$ exceeds the critical point
\begin{eqnarray}
A_{s}^{crit}=\sqrt{\frac{8}{3}m^{2}}
\end{eqnarray}
where $m^{2}=m_{1}^{2}+m_{2}^{2}+m_{S}^{2}$.
As this formula indicates, when $m^{2}>0$, $A_{s}^{crit}$ exists, and 
passage of the potential from small to large trilinear coupling regime 
is a first order phase transition, namely, all VEV's are discontinuous 
at $A_{s}^{crit}$. On the other hand, when $m^{2}<0$, there is no 
critical point at all; transition is exclusively second order 
\cite{lang,zprime}. However, independent of the sign of $m^{2}$ and the order 
of the transition, in the limit of large $A_{s}$, all VEV's converge the 
solution 
\begin{eqnarray}
v_{1}\sim v_{2}\sim v_{s} \sim \frac{A_{s}}{\sqrt{2} h_{s}}
\end{eqnarray}
with which one can fix $A_{s}=h_{s} (2 G_{F})^{-1/2}$ by using the $W^{\pm}$ mass. 
When the potential (3) possesses a HTCDM, in the basis $(Re[H_{1}^{0}], 
Re[H_{2}^{0}], Re[S^{0}])^{T}$, CP-even Higgs mass-squared matrix is 
given by 
\begin{eqnarray}
({\cal{M}}^{2})_{h}=(G_{F}^{-1}/4)\left (\begin{array}{c c c}
2\lambda_{1}+h_{s}^{2}&\lambda_{12}-h_{s}^{2}&\lambda_{1S}-h_{s}^{2}\\\ 
\lambda_{12}-h_{s}^{2}&2\lambda_{2}+h_{s}^{2}&\lambda_{2S}-h_{s}^{2}\\\
\lambda_{1S}-h_{s}^{2}&\lambda_{2S}-h_{s}^{2}&2\lambda_{S}+h_{s}^{2}
\end{array}\right)
\end{eqnarray}
where $\lambda_{1}=\lambda_{2}$ and $\lambda_{1S}=\lambda_{2S}$, because 
$Q_1=Q_2$ as assumed above, when discussing HTCDM. Diagonalization of 
the CP-even Higgs mass-squared matrix gives the following mass spectrum 
\begin{eqnarray}
m_{h_{1}}=(G_{F}^{-1/2}/2)h_{s}\;\;,\; 
m_{h_{2}}=(G_{F}^{-1/2}/2)\sqrt{h_{s}^{2}+G^2/2}\;\;,\; 
m_{h_{3}}=(G_{F}^{-1/2}/2)\sqrt{6g_{Y'}^{2}+h_{s}^2}
\end{eqnarray}
in the increasing order. The physical mass eigenstates 
$(h_1, h_2, h_3)^{T}$ are related to the basis vector $(Re[H_{1}^{0}],  
Re[H_{2}^{0}], Re[S^{0}])^{T}$ via the diagonalizing matrix 
\begin{eqnarray}
{\cal{R}}=\left (\begin{array}{c c c}
1/\sqrt{3}&1/\sqrt{3}&1/\sqrt{3}\\\
-1/\sqrt{2}&1/\sqrt{2}&0\\\
-1/\sqrt{6}&-1/\sqrt{6}&2/\sqrt{6}
\end{array}\right)
\end{eqnarray} 
Having matrix ${\cal{R}}$ at hand, one can easily compute the coupling 
strengths of $h_{k}$ to $Z_{i}Z_{j}$ for $i,j=1,2$ and $k=1,2,3$. In 
fact, Table III gives a list of these couplings for all possible cases. 
As one notices this table has important implications for the mechanism 
of Higgs production through the Higgs-strahlung type processes. First of 
all, the sum of the squared $h_{k}Z_{1}Z_{1}$ couplings equals the square of 
the SM  $H\,Z\,Z$ coupling. Thus, just like MSSM, mixings in the scalar 
sector result in a reduction of the coupling strength, implying a smaller 
production cross section than that of the SM.
 
Next, one observes that some couplings vanish. While $h_{1}$ and $h_{3}$ 
have only diagonal couplings in the form $Z_{i}Z_{i}$, $h_{2}$ has only 
$Z_{1}Z_{2}$ type coupling. Hence, in a transition of the form 
$Z_{i}\rightarrow h_{k} Z_{j}$, if the initial and final vector bosons are 
identical then only lightest and heaviest Higgs bosons could be radiated 
off. Unlike this, if vector bosons are not identical, then the radiated 
scalar can only be next-to-lightest Higgs. This well defined spectrum 
of the produced scalars may be important in a particular collider search, 
say LEP2 or NLC. This completes the discussion of the scalar sector of 
the $Z^{\prime}$ models at the tree level. When we discuss the loop 
effects we shall reanalyze some quantities derived in this section.
\subsection{Relevant minimum in NMSSM}
Having discussed the scalar sector of the $Z^{\prime}$ models, we now 
start analyzing NMSSM scalar potential (3) in reference to Table I. As 
mentioned at the beginning of this section, we require supersymmetry be 
broken around the weak scale,  and thus, Higgs VEV's are of the 
same order of magnitude. Restrictions on $Z^{\prime}$ models do not have 
any analogue in NMSSM, and one is generally free to realize any kind of 
minimum as long as VEV's do not have too big splittings among them. In 
fact, all relevant portions of the NMSSM parameter space have already been 
analyzed in \cite{nmssm} whose results will not be reproduced here. 
For a comparative and parallel study of the $Z^{\prime}$ models and 
NMSSM, it would be 
convenient to discuss the latter one in that portion of the parameter 
space required by the former as it is severely constrained by the LEP 
Z-pole data. Thus, differently than \cite{nmssm}, we shall discuss 
NMSSM 
also in the limit of large Higgs trilinear coupling. Actually, NMSSM 
scalar potential (3) would possess a HTCDM provided that 
$|h_{s}A_{s}+k_{s}A_{k}|$ exceeds the critical value
\begin{eqnarray}
\tilde{A}_{s}^{crit}=\sqrt{\frac{8}{9}\lambda m^{2}}
\end{eqnarray}
where $\lambda=3h_{s}^{2}+k_{s}^{2}+2h_{s}k_{s}$, and 
$m^{2}=m_{1}^{2}+m_{2}^{2}+m_{S}^{2}$. In NMSSM, instead of 
$|h_{s}A_{s}|$, one has $|h_{s}A_{s}+k_{s}A_{k}|$ characterizing the 
minimum of the potential. At this point one should bare in mind that 
$S^{3}$ coupling $|k_{s}A_{k}|$ singles out $S$ and its large values 
automatically creates $v_{s}> > v_{1,2}$, rather than  $v_{1}\sim v_{2}\sim 
v_{s}$. Consequently, if one wishes to obtain a HTCDM for the potential 
(3), $|k_{s}A_{k}|$ must be much less than $|h_{s}A_{s}|$ so that 
$\tilde{A}_{s}^{crit}$ approximately applies to $A_{s}^{crit}$. This is 
an approximate statement because one cannot ignore $|k_{s}A_{k}|$ 
completely due to the axion problem mentioned in the Introduction. Just 
like the $Z^{\prime}$ models, type of the transition is sensitive to the 
sign of $m^{2}$, however, independent of this, for large enough 
$|h_{s}A_{s}|$, all VEV's converge to the same value given by 
\begin{eqnarray}
v_{1}\sim v_{2}\sim v_{s}\sim 
\frac{3}{\lambda}\frac{h_{s}A_{s}+k_{s}A_{k}}{\sqrt{2}}
\end{eqnarray}
which mainly follows $h_{s}A_{s}$ since $|k_{s}A_{k}|< <|h_{s}A_{s}|$. 
An analysis of the $Z^{\prime}$ models reveals that CP-even Higgs 
mass-squared matrix is highly sensitive to gauge and Yukawa couplings as 
can be seen from (14). In NMSSM, the prescription $|k_{s}A_{k}|< 
<|h_{s}A_{s}|$ implies two distinct cases to be analyzed in detail:
\begin{eqnarray}
|k_{s}A_{k}|< <|h_{s}A_{s}| \Longrightarrow \left\{ \begin{array}{c}
k_{s}\approx h_{s}\;\; and\;\; |A_{k}| < < |A_{s}|\Longrightarrow \mbox{NMSSM1}\\
k_{s}< < h_{s}\;\; and\;\; |A_{k}| \approx |A_{s}| \Longrightarrow \mbox{NMSSM2}
\end{array}\right.
\end{eqnarray}
where we named the two cases as NMSSM1 and NMSSM2 for later use.
It is convenient to discuss the implications of these two cases seperately.
$$
\underline{\mbox{NMSSM1}}:
$$
In this case one can replace $k_s$ by $h_{s}$ so that $\lambda\approx 6
h_{s}^{2}$. Furthermore, neglecting $A_{k}$ in comparison with $A_s$, one 
obtains   
\begin{eqnarray}
A_{s}^{crit}\approx \sqrt{2}\times \sqrt{\frac{8}{3} m^{2}}
\end{eqnarray}
which, when sufficiently exceeded by $A_{s}$, implies the VEV's
\begin{eqnarray}
v_{1}\sim v_{2} \sim v_{s} \sim \frac{1}{2}\times \frac{A_{s}}{h_{s}\sqrt{2}}
\end{eqnarray}
With these VEV's, in the basis  $(Re[H_{1}^{0}], Re[H_{2}^{0}], 
Re[S^{0}])^{T}$, mass-squared matrix for CP-even scalars turns out to be 
\begin{eqnarray}
({\cal{M}}^{2})_{h}=(G_{F}^{-1}/4)\left (\begin{array}{c c c}
G^2/4+3 h_{s}^{2}/2&-G^2/4+h_{s}^{2}/2&0\\\
-G^2/4+h_{s}^{2}/2&G^2/4+3 h_{s}^{2}/2&0\\\
0&0&4h_{s}^{2}
\end{array}\right)
\end{eqnarray}
the diagonalization of which yields Higgs spectrum with masses 
\begin{eqnarray}
m_{h_{1}}=(G_{F}^{-1/2}/2\sqrt{2})\sqrt{h_{s}^{2}+G^{2}}\;\;,\;
m_{h_{2}}=(G_{F}^{-1/2}/\sqrt{2})h_{s}\;\;,\;
m_{h_{3}}=(G_{F}^{-1/2})h_{s}
\end{eqnarray}
in the increasing order. The physical mass eigenstates
$(h_1, h_2, h_3)^{T}$ are related to the basis vector $(Re[H_{1}^{0}],
Re[H_{2}^{0}], Re[S^{0}])^{T}$ via the diagonalizing matrix
\begin{eqnarray}
{\cal{R}}=\left (\begin{array}{c c c}
-1/\sqrt{2}&1/\sqrt{2}&0\\\
1/\sqrt{2}&1/\sqrt{2}&0\\\
0&0&1
\end{array}\right)
\end{eqnarray}
As this diagonalizing matrix shows, the heaviest Higgs gets contributions 
only from $Re[S^{0}]$, and the lighter Higgs particles get contributions 
only from the neutral CP-even components of $H_{1}$ and $H_{2}$. In this
sense doublets and the singlet decouple, and produce their Higgs spectra.
One recalls that the situation 
in $Z^{\prime}$ models was different; there it was only the next-to-lightest 
Higgs that was independent of $Re[S^{0}]$. This forms a clear distinction 
between the two models. With ${\cal{R}}$ matrix at hand, it is easy to 
compute the strength of the coupling between a $Z$ line and a 
CP-even Higgs as already listed in the first column of Table IV. As the table 
shows only next-to-lightest Higgs is radiated off a $Z$ line, and the 
lightest and the heaviest Higgs scalars cannot be produced. In this way 
one concludes that Higgs-strahlung type processes can lead to the production of 
next-to-lightest Higgs only. Moreover, in terms of $Z_1$ 
couplings, $Z^{\prime}$ models and NMSSM are complementary to each other. 
This completes the discussion of the NMSSM1 in terms of 
its particle spectrum and implications for Bjorken production of the Higgs 
particles.
$$
\underline{\mbox{NMSSM2}}:
$$
In this case one can replace $A_{k}$ by $A_{s}$ and neglect $k_{s}$ in 
comparison with $h_s$ so that $\lambda\approx 3 h_{s}^{2}$. Then, in 
exact similarity with the $Z^{\prime}$ models, one gets 
\begin{eqnarray}
A_{s}^{crit}\approx \sqrt{\frac{8}{3} m^{2}}
\end{eqnarray}
which, when sufficiently exceeded by $A_{s}$, implies the VEV's
\begin{eqnarray}
v_{1}\sim v_{2} \sim v_{s} \sim \frac{A_{s}}{h_{s}\sqrt{2}}
\end{eqnarray}
With these VEV's, in the basis  $(Re[H_{1}^{0}], Re[H_{2}^{0}],
Re[S^{0}])^{T}$, mass-squared matrix for CP-even scalars turns out to be 
\begin{eqnarray}
({\cal{M}}^{2})_{h}=(G_{F}^{-1}/4)\left (\begin{array}{c c c}
G^2/4+h_{s}^{2}&-G^2/4&0\\\
-G^2/4&G^2/4+h_{s}^{2}&0\\\
0&0&h_{s}^{2}
\end{array}\right)
\end{eqnarray}
the diagonalization of which yields the particle spectrum
\begin{eqnarray}
m_{h_{1}}=(G_{F}^{-1/2}/2)h_{s}\;\;,\;
m_{h_{2}}=(G_{F}^{-1/2}/2)\sqrt{h_{s}^{2}+2k_{s}^{2}}\;\;,\;
m_{h_{3}}=(G_{F}^{-1/2}/2)\sqrt{h_{s}^{2}+G^{2}/2}\;\;,\;
\end{eqnarray}
in the increasing order. The physical mass eigenstates
$(h_1, h_2, h_3)^{T}$ are related to the basis vector $(Re[H_{1}^{0}],
Re[H_{2}^{0}], Re[S^{0}])^{T}$ via the diagonalizing matrix
\begin{eqnarray}
{\cal{R}}=\left (\begin{array}{c c c}
1/\sqrt{2}&1/\sqrt{2}&0\\\
0&0&1\\\
-1/\sqrt{2}&1/\sqrt{2}&0
\end{array}\right)
\end{eqnarray}
In comparison with NMSSM1, here one encounters some novel 
aspects. Though the CP-even Higgs mass-squared matrices (27) and (22) have the 
same form their dependences on $h_{s}^{2}$ are not identical because of the fact
that $k_{s}$ plays different roles in two cases. However, one still expects doublets
and the singlet to decouple due to the form of the mass-squared matrix (27).
Indeed, as the diagonalizing matrix (29) shows this time it is the next-to-lightest 
Higgs that is a pure singlet state as opposed to NMSSM1 where the heaviest Higgs
was a pure singlet. One notices that since $k_{s}<<h_{s}$, $h_{1}$ 
and $h_{2}$ are nearly degenerate in mass. A comparison of the 
diagonalizing matrices (24) and (29) implies the cyclic interchange 
$h_{3}\longleftrightarrow h_{2}$, $h_{2}\longleftrightarrow h_{1}$ , and 
$h_{1}\longleftrightarrow h_{3}$. Thus, in NMSSM2 $h_2$ is a pure 
singlet, and the heaviest and the lightest Higgs particles get 
contribution from only the neutral CP-even parts of the Higgs doublets. 
In this case, Higgs-strahlung type processes 
support the production of the lightest and the heaviest Higgs scalars 
only. With the diagonalizing matrix (29), one can compute the couplings 
of Higgs scalars to a $Z$ line. In fact, the second column of Table IV 
indicates these couplings. As we see, only the lightest Higgs $h_1$ can 
be produced in a Higgs-strahlung type process in NMSSM2.
The heavier Higgs scalars cannot be obtained in such processes.   
\subsection{Effects of the Radiative Corrections} 
The discussions above show that, in the limit of large Higgs trilinear couplings, 
both $Z^{\prime}$ models and NMSSM approach a HTCDM in which 
all vacuum expectation values are of the same order. However, for a proper analysis 
of the particle spectrum it is necessary to investigate the effects of the
radiative corrections on the stability of the HTCDM. Effects of the radiative 
corrections have already been discussed for $Z^{\prime}$ models in \cite{dur}, and 
NMSSM in \cite{nmssm-people}. The radiative effects of the entire particle spectrum
on the Higgs potential could be parametrized by using the effective potential 
approximation. To one-loop accuracy the effective potential has the
 usual Coleman-Weinberg form
\begin{eqnarray}
V_1=V+\frac{1}{64\pi^2}
Str{\cal{M}}^{4}\ln\frac{{\cal{M}}^{2}}{Q^2}
\end{eqnarray}
where ${\cal{M}}$ is the Higgs field dependent mass-matrices of the fields 
entering the supertrace $Str=(-1)^{2J}(2J+1)Tr$. Here $Q$ is the renormalization 
scale. Generally, unless supersymmetry is broken below $\sim$ TeV the loop 
effects as well as the direct production of the supersymmetric particles
are hard to detect in present and near future experiments. However, then 
the logarithms in the Coleman-Weinberg effective potential may not be satisfactorily 
large. In spite of this fact, this formula is still satisfactory for large 
enough Yukawa couplings (See the second reference in \cite{mssm}). Among all supersymmetric
particle spectrum especially top quark and top squarks are important due to 
the large value of top Yukawa coupling, $h_{t}\sim \sqrt{2}$. Then, taking only the dominant 
stop and top quark contributions into account, among all potential parameters 
only the following ones get significantly corrected:
\begin{eqnarray}
\hat{A}_{s}&=&A_{s}+\beta_{h_{t}}S_{\tilde{t}Q}A_{t}\nonumber\\
\hat{m}_{2}^{2}&=&m_{2}^{2}+\beta_{h_{t}}\Big[(A_{t}^{2}+A^{2})S_{\tilde{t}Q} 
-A^{2}\Big]\nonumber\\ 
\hat{\lambda}_{1s}&=&\lambda_{1s}+\beta_{h_{t}} S_{\tilde{t}Q} h_{s}^{2}\\
\hat{\lambda}_{2}&=&\lambda_{2}+\beta_{h_{t}}S_{\tilde{t}t}h_{t}^{2}\nonumber
\end{eqnarray}  
with
\begin{eqnarray}
\beta_{h_{t}}&=&\frac{3}{(4\pi)^{2}}h_{t}^{2}\nonumber\\
S_{\tilde{t}Q}&=&\ln \frac{m_{\tilde{t}_{1}}m_{\tilde{t}_{2}}}{Q^{2}}\\
S_{\tilde{t}t}&=&\ln 
\frac{m_{\tilde{t}_{1}}m_{\tilde{t}_{2}}}{m_{t}^{2}}\nonumber
\end{eqnarray}
where the last two quantities represent the splitting between 
stops and the scale $Q$, and top quark mass, respectively. $A_{t}$ is the top 
trilinear coupling coming from the superpotential (1) or (2), and the 
the remaining mass parameter $A^{2}$ is the sum of the soft mass-squareds of top 
squarks; $A^{2}=m_{\tilde{Q}}^{2}+m_{\tilde{U}}^{2}$. In deriving these 
one-loop corrections we assumed $m_{\tilde{Q}}^{2}\sim m_{\tilde{U}}^{2}$ 
in accordance with the FCNC constraints \cite{fcnc}, and expanded the 
stationarity conditions in powers of stop splitting 
$m_{\tilde{t}_{2}}^{2}- m_{\tilde{t}_{1}}^{2}$. Detailed expressions  
for stop mass-squared matrix can be found in \cite{dur,nmssm-people}.
 
One recalls from the discussions of the tree-level potential that the 
existence of the Higgs trilinear coupling driven minimum of the potential 
can be characterized by the threshold value $A_{s}^{crit}$ of $A_s$. This 
threshold value of $A_s$ is highly precise if the sum of the soft 
mass-squareds of the Higgs fields are positive, and irrespective of the 
order of the transition potential possesses a HTCDM if $A_s$ is 
sufficiently large compared to $A_{s}^{crit}$. Thus, it is convenient to 
analyze the effects of the radiative corrections on the  vacuum structure 
starting with a HTCDM at the tree-level. If ${\cal{A}}_{s}^{crit}$ is 
the critical value of $A_s$ in the presence of the radiative corrections,
one has
\begin{eqnarray}
{\cal{A}}_{s}^{crit}=\frac{A_{s}^{crit}}{1+\delta}
\end{eqnarray}
where $\delta$ represents the effects of the radiative corrections, and 
can be expressed solely in terms of the Higgs- and top-trilinear 
couplings and the Yukawa couplings:
\begin{eqnarray}
\delta&=&\frac{1}{6}\beta_{h_{t}} S_{\tilde{t}Q}\Big[ - 2 +12 \frac{A_{t}}{A_{s}}-16
\left(\frac{A_{t}}{A_{s}^{crit}}\right)^{2}\Big]\nonumber\\&+&
\frac{1}{18} \beta_{h_{t}}(1-S_{\tilde{t}Q})\Big[16
\left(\frac{A_{t}^{crit}}{A_{s}^{crit}}\right)^{2}+3 \left(\frac{A_{t}}{A_{s}^{crit}}\right)^{2}\Big]\nonumber\\&-&
\frac{1}{3}\beta_{h_{t}} S_{\tilde{t}t}\left(\frac{h_{t}}{h_{s}}\right)^{2} 
\end{eqnarray}
where, following the discussion of CCB minima in MSSM in \cite{casas}, we 
introduced the critical value of $A_{t}$ via the relation
\begin{eqnarray}
{A_{t}^{crit}}^{2}+3\mu_{s}^{2}=m_{\tilde{Q}}^{2}+m_{\tilde{U}}^{2}
\end{eqnarray}
which represents the threshold value of $A_{t}$ for which color and/or 
charge breaking just starts taking place, for a given 
$\mu_{s}=\frac{h_{s}v_{s}}{\sqrt{2}}$. In these formuale, $A_s\geq 
A_{s}^{crit}$ and $A_{t}\leq A_{t}^{crit}$ if the vacuum state under 
concern is a non-CCB HTCDM. The exact value of $\delta$ requires a full 
specification of the scalar sector of the theory including the squark 
trilinear mass terms in superpotentials (1) or (2). The criterium induced 
by $\delta$ has a severe dependence on its sign: If $\delta <0$, 
${\cal{A}}_{s}^{crit}>A_{s}^{crit}$, and thus tree-level HTCDM might 
disappear if the actual value of $A_s$ were close to $A_{s}^{crit}$ at 
the tree level. In the opposite case of $\delta >0$, 
${\cal{A}}_{s}^{crit}<A_{s}^{crit}$ and HTCDM is supported by the 
radiative corrections. Since the Higgs VEV's are of the same order, 
satisfying the CDF value of the top-quark requires $h_{t}\sim \sqrt{2}$, 
so that effects of the radiative corrections are no way negligable. 
Moreover, for the reasonable set of parameters, $h_{t}\sim \sqrt{2}, 
\;A_{t}^{crit}\sim A_{s}^{crit},\; Q^{2}\sim m_{t}^{2},\; 
m_{\tilde{t}_{1}}m_{\tilde{t}_{2}}\sim G_{F}^{-1}$ and $h_{s}\sim G$, 
$\delta$ turns out to be  negative so that radiative corrections 
prefer ${\cal{A}}_{s}^{crit}>A_{s}^{crit}$ which can destabilize the 
vacuum state unless $A_{s} >>A_{s}^{crit}$ at the tree level. Radiative 
corrections modify mass matrices of all scalars, in particular that of the 
CP-even Higgs scalars by addition of the following matrix 
\begin{eqnarray}
(\delta{\cal{M}}^{2})_{h}=\beta_{h_{t}}S_{\tilde{t}t}\left 
(\begin{array}{c c c} \mu_{s}A_{t}&-\mu_{s}A_{t}&\mu_{s}(2\mu_{s}-A_{t})\\\
-\mu_{s}A_{t}&4m_{t}^{2}&-\mu_{s}A_{t}\\\
\mu_{s}(2\mu_{s}-A_{t})&-\mu_{s}A_{t}&\mu_{s}A_{t}
\end{array}\right)
\end{eqnarray}
which is obtained by assuming $S_{\tilde{t}Q}\sim S_{\tilde{t}t}$, and
a small stop splitting. In fact, one can write $S_{\tilde{t}t}=\ln{
(1+(3/2)(h_{s}^{2}/h_{t}^{2})+{A_{t}^{crit}}^{2}/2m_{t}^{2})}$ using the 
definition of $A_{t}^{crit}$ (35). As $Q^{2}\sim m_{t}^{2}$ is the natural 
renormalization scale, this approximation is reasonable. The entries of the 
tree-level mass matrices (14) of $Z^{\prime}$ models, and (22) and (27) 
of NMSSM are now modified with the addition of (36). Let us note an 
important difference between $Z^{\prime}$ models and NMSSM in terms of 
the behaviour of their mass matrices under radiative corrections. 
In $Z^{\prime}$ models, (14), with its all elements are non-vanishing, gets
non-vanishing contributions to each of its elements through the radiative 
corrections, preserving its form. On the other hand, in NMSSM, tree-level 
mass matrices, (22) and (27), already have some of their elements 
vanishing so that there is no mixing between the doublet and 
singlet contributions. However, with the addition of the radiative 
corrections (36), this simple tree-level picture is destroyed, and now 
singlet and doublet sectors do mix. Even in the limit of small $A_{t}$, 
$Re[H_{1}^{0}]Re[S^{0}]$ type mixing cannot be avoided. 

If $A_{s}$ is sufficiently large compared to the radiatively 
corrected critical point ${\cal{A}}_{s}$ (33) then potential (3) will 
definitely have a HTCDM in which the VEV's are proportional to $A_{s}$ 
and their values are eventually fixed by the Fermi scale $G_{F}^{-1}$. Thus 
one can safely take VEV's equal in (36). Factoring out $G_{F}^{-1}/4$ 
from (36), one observes that $\mu_{s}^{2}$ and $m_{t}^{2}$ terms 
contribute by $0.08\, h_{s}^{2} S_{\tilde{t}t}$ and $0.3 
S_{\tilde{t}t}$, respectively. Thus, especially the top quark 
contribution is important. Although one can 
analytically obtain the effects of (36) on the mass spectrum and 
couplings, the results will be algebraically involved; and thus, we will 
numerically analyze the consequences of radiative corrections in the next 
section.
\section{Higgs search via Bjorken processes}
The search for the Higgs particles is one of the most important issues at 
LEP2 and NLC. Though two-fermion processes are still of interest, the true 
novelty at high energy $e^{+}e^{-}$ colliders will come from four-fermion 
processes, among which the $s$-channel Higgs-strahlung process 
$e^{+}e^{-}\rightarrow Z H$ is the most important one for Higgs 
production. In fact, other processes in 
which Higgs is formed in $WW$ and $ZZ$ $t$- channel collisions have 
smaller cross sections at LEP2 energies, and their interference with the 
Higgs- strahlung type processes could be avoided by preventing $e$ and 
$\nu_{e}$ from the final products \cite{LEP}. Above all, we should
analyze four-fermion (4f) processes because intermediate state Higgs 
particles show up as resonances when their mass and the invariant mass 
flow to that channel coincide. In the much simpler process 
$e^{+}e^{-}\rightarrow Z H$ one cannot trigger the multi-Higgs structure 
of the models under concern as the products in a specific collision 
process are fixed. With these constraints in mind, we analyze the 
following four-fermion process 
\begin{eqnarray}
e^{+}e^{-}\rightarrow Z_{i}\rightarrow (Z_{j}\rightarrow 
\bar{f}_{2}f_{2})\;\; (h_{k}\rightarrow \bar{f}_{1}f_{1})
\end{eqnarray}  
where, depending on the $e^{+}e^{-}$ CM energy $\sqrt{s}$, vector bosons 
and Higgs bosons above may come to the physical shell. Computation of the 
cross section for 4f processes like this is a highly complicated 
problem due to the phase space integration \cite{benimkiler}, and one mostly resorts to 
numerical techniques \cite{LEP}. Despite this, however,
we can extract the necessary information about the Higgs structure of the 
underlying models without performing a full calculation if we take the 
ratio of the differential cross section to that of the SM. Before going 
into the details of such a calculation it is convenient to describe the 
properties of $Z^{\prime}$ models due to its complicated Higgs-vector 
boson sector. For $Z^{\prime}$ models, in HTCDM, $Z-Z^{\prime}$ mixing 
angle (9) is vanishingly small, and we assume it remains small also when 
radiative corrections are included.  Assuming further $Z^{\prime}$ be 
leptophobic, it is seen that $Z_{2}$ does not contribute to (37), so an 
analysis of (37) with $Z_1\equiv Z$ is sufficient. 
  
To calculate the total cross section $\sigma$ one needs to integrate over 
the phase space of the final particles, which is not possible by hand 
\cite{LEP}. Moreover, in the case of polarized electron-positron beams 
task will be much more complicated. The amplitude for the scattering process 
(37) in the models under concern differs from that in the SM by the 
multitude of Higgs scalars reflecting them in $Z Z h_{k}$ and $h_{k} 
\bar{f}_{1}f_{1}$ couplings. This observation greatly simplifies the 
analysis if one integrates over first $\bar{f}_1$ and $f_1$ phase spaces 
using the momentum conservation, and identifies the remaining task to be 
done with that of the SM process. Then the ratio of the resulting cross section 
to that of the SM depends only on the invariant mass $p^{2}$ of the 
$\bar{f}_{1}f_{1}$ system: 
\begin{eqnarray}
\frac{d\sigma_{Z^{\prime},\; NMSSM} (p^{2})}{d\sigma_{SM} (p^{2})} \equiv 
 I(p^{2})= \mid \sum_{k} R_{Z Z h_{k}} {\cal{R}}_{a k} 
\frac{p^{2}-m_{H}^{2}-im_{H}\Gamma_{SM}}{p^{2}-m_{h_{k}}^{2}-im_{h_{k}} 
\Gamma_{k}}\mid ^{2}
\end{eqnarray}
where $R_{Z Z h_{k}}$ are $Z Z h_{k}$ couplings, listed in Tables III and 
IV, in units of corresponding SM coupling $G/2$. Higgs-fermion 
couplings are parametrized by $a$, which can be 1 and 2 for down- and 
up-type fermions, respectively. Furthermore, $\Gamma_{k}$ and $\Gamma_{SM}$ 
designate the total widths of $h_{k}$ and $H$, respectively. $p^{2}$, 
invariant mass flowing into the Higgs branch, has the kinematical range 
of  $4m_{f_{1}}^{2}$ to 
$(\sqrt{s}-2m_{f_{2}})^{2}$. As $p^{2}$ varies in this range  
$I(p^{2})$ will be sharply peaked at each $m_{h_{k}}$, as long as 
$m_{h_{k}}$ is kinematically accessible. Finally, one notices that 
$I(p^{2})$ is independent of the beam polarizations. 
 
For the compactness of the presentation, we introduce the dimensionless 
parameter $x^2=p^2/M_{H}^{2}$ which equals unity when the SM Higgs mass 
resonance is encountered. In what follows, we take $M_{H}$ equal to the  
lightest Higgs boson mass in the  particular model under concern. In this 
sense, at $x=1$ lightest Higgs resonance occurs.  As $x$ gets higher and 
higher values other Higgs scalars with non-vanishing couplings will be 
excited in the order of increasing mass. For trilinear Higgs Yukawa 
coupling $h_s$, we assume $h_{s}\approx G$. Actually, $h_s$ can be choosen 
as low as $h_{s}^{min}\approx 0.36$ at which the present LEP lower
bound of the lightest Higgs mass is exceeded at the tree level.
For $h_{s}\sim G$, one gets $\mu_{s} \approx M_{Z}$, which is a reasonable 
scale. On the other hand, for $g_{Y'}$ we assume the usual GUT 
constraint of $g_{Y'}\sim \sqrt{5/3} g_{Y}$. The $S^{3}$ Yukawa coupling 
$k_{s}$ is either $\sim h_{s}$ (NMSSM1), or $<< h_{s}$ 
(NMSSM2). For $|k_{s}A_{k}| << |h_{s}A_{s}|$, as is 
necessary for obtaining HTCDM, $A_{k}$ dependence of the masses and the 
couplings cancel, as was illustrated before. In addition to these, we 
take $|A_{t}|<<\mu_{s}$, so that $A_{t}$ dependence of the radiative 
corrections to the CP-even Higgs mass-squared matrix (36) can be 
neglected. Finally, we approximate $\Gamma_{SM}$ by $\Gamma(H\rightarrow 
\bar{b}b)+\Gamma(H\rightarrow \bar{c}c)$, and assume all scalars have the 
same width. This last assumption is falsified especially when the Higgs mass 
exceeds $M_{Z}$, but it does not affect the main conclusion of the work 
because the width of a particular resonance is not of central concern. At the 
final state we take $f_1=b$ and $f_2=\mu$, which is convenient for 
detection purposes at LEP2 \cite{LEP}.
 
With the above mentioned values of parameters, one can calculate all 
couplings and masses. In fact, tree- and loop- level values of the 
lightest Higgs mass are tabulated in Table V, for $Z^{\prime}$ models, 
and NMSSM for both cases in (19). An analysis of $I(x)$ (38) reveals that 
it is necessary to have $\sqrt{s}\geq 350$ GeV for all Higgs scalars be 
excited in both models. This value of $\sqrt{s}$ is pretty much above the 
maximum value of $205$ GeV aimed at LEP2, and thus one is to wait for NLC 
operation for the experimental realization of these scalar spectra, if 
any. 

In Fig. 1 we show $I(x)$ for $Z^{\prime}$ models, at $\sqrt{s}=350$ GeV, 
with and without the radiative corrections. Here dashed curve represents 
the tree-level analysis, and as expected there is a single resonance 
curve at  $x\approx 2.1$ corresponding to the heaviest Higgs whose 
coupling is given in Table II. In this graph, the full curve shows $I(x)$ when 
radiative corrections are included. Under radiative corrections none of 
the couplings remains vanishing, and therefore, effects of the 
scalars which were inert at the tree level show up. In this sense, the first 
resonance curve at $x\approx 1.2$ corresponds to the next-to-lightest 
Higgs which was absent in the tree- level $I(x)$. That this resonance is 
much narrower than that of the heaviest Higgs located at $x\approx 2$ is 
caused by the smallness of $R_{Z\,Z\,h_{2}}$. Both tree- and loop-level 
$I(x)$ has a non-negligable value at $x=1$ because of the fact that 
lightest Higgs is observable at the tree-level and has still a large 
enough coupling at the loop level.

In Fig. 2 we present $I(x)$ for  NMSSM1 described in equation (19). At the 
tree-level, in accordance with Table IV, only next-to-lightest Higgs is 
observable as is evidenced by the resonance at $x\approx 1.6$. Both 
dashed and full curves have vanishingly small values at $x\approx 1$ 
due to the fact that $R_{Z\,Z\,h_{1}}$, which vanishes at the tree-level, is still 
small compared to  $R_{Z\,Z\,h_{2}}$. When radiative corrections are 
included place of the next-to-lightest Higgs resonance practically 
remains the same, and there is a tiny resonance curve at $x\sim 1.83$ 
representing the heaviest Higgs contribution. Such a narrow resonance is 
caused by the smallness of $R_{Z\,Z\,h_{3}}$ compared to $R_{Z\,Z\,h_{2}}$. 
 
In Fig. 3 $I(x)$ for NMSSM2 is depicted. At the 
tree-level, in accordance with Table IV, only the lightest Higgs observable 
so that the dashed line $I(x)\approx 0.5$ arises. Differently than the 
models analyzed in Figs. 1 and 2 in this model radiative corrections cause 
important modifications in the spectrum. This is mainly caused by the 
smallness of the  elements of the mass-squared  matrix compared to 
the two cases discussed above which make them more sensitive to the 
radiative corrections. Consequently, when the radiative corrections 
are included next-to-lightest Higgs is seen to have dominant couplings 
compared to others, so that aymptotics of the full curve are entirely 
determined by $h_{2}$. In fact, $I(x)$ has vanishingly small values at 
$x\approx 1$ and $x\approx 1.4$ due to this reason. The resonance curves of 
$h_{2}$ and $h_{3}$ are close to each other, narrower than tree-level 
ones of the previous cases, and located at $x\approx 1.1$ and $x\approx 
1.3$, respectively.

\section{Conclusions and Discussions}
The analysis carried out in this work primarily concentrates on the
similarities and dissimilarities between the $Z^{\prime}$ models and NMSSM 
concerning their neutral Higgs scalars. The two models have been comparatively
analyzed in that minimum of the potential characterized by the large values
of trilinear soft mass coupling the SM singlet and the Higgs doublets. The issue 
of radiative corrections deals essentially with the stability of this minimum 
together with the corresponding Higgs mass--squared matrix. The radiative corrections
are conveniently parametrized using the effective potential approximation. As a 
case study we analyzed Higgs production via Bjorken process at an $e^{+}e^{-}$
collider. Both models are comparatively analyzed referring basically to the 
variation of the cross section with the di-fermion invariant mass. 

Once a high energy collider is given there appear several scattering 
processes whose signature necessarily depend on the details of the underlying model. 
This becomes especially clear after a comparative study of the SM and MSSM, for
example, at $e^{+}e^{-}$ linear colliders \cite{review}. On the other hand, characteristic 
to all supersymmetric models, there are sfermions, charginos, neutralinos, 
and fermions in addition to the Higgs particles whose distinguishability needs be studied.
The two minimal extensions of the MSSM, that is, $Z^{\prime}$ models and NMSSM, are already 
already distinguishable from the MSSM and 2HDM in terms of their particle spectrum. Moreover, there 
are also differences between $Z^{\prime}$ models and NMSSM as they have  different
number of neutral gauge bosons (two for the former, one for the latter), psedoscalars
(one for the former, two for the latter), and neutral gauginos (three for the former, 
two for the latter). Therefore, for processes with such
signature, it can be easy to know what model is the underlying one. For example, 
the neutralino pair production,
$e^{+}e^{-}\rightarrow \chi_{i}^{0}\chi_{j}^{0}$ or associated Higgs production 
$e^{+}e^{-}\rightarrow H_{i}P_{i}$ ($P$ is a pseudoscalar) are candidate processes.
Despite such differences, however, the two models have equal number of 
CP--even Higgs particles. The primary goal of the future colliders is the 
discovery of the Higgs particle whose most likely production mode is the
Bjorken process. That this process does not involve the pair production 
of the supersymmetric particles is especially appealing as it could 
be realized for moderate collider energies. Due to the reasons mentioned above, 
a discussion of the distinguishability of the CP--even Higgs scalars of the
$Z^{\prime}$ models and NMSSM is an important issue by itself. We have concentrated 
here on HTCDM (where $Z^{\prime}$ models are likely to have for satisfying the LEP 
constraints) in which the Higgs potentials of the two model are similar apart 
from their respective parameters. The results of the analysis are presented in the
tabels and figures for both tree--level and one--loop Higgs potentials. NMSSM, having 
no gauge extension at all, is particularly interesting in HTCDM. For example, a glance 
at Table IV shows that these couplings are identical to those of the minimal model. 
However, one--loop $R_{ZZh_{i}}$ couplings for both NMSSM1 and NMSSM2 destroy this tree--level
picture by causing other coupligs to have non--vanishing values. Then, as suggested by 
Figs. 2 and 3 there appear new resonances allowing, eventually, for distinguishability. 
This examplifying case particularly shows how important the radiative corrections are. 
At the tree level, there is a certain hierarchy, in particular, some couplings vanish 
identically as shown in Tables 3 and 4. However, the radiative corrections are strong 
enough to elevate other couplings to non-vanishing values. This is particularly observed
in the figures (solid curves) where there arise additional resonances compared to 
the tree--level results (dashed curve) corresponding to additional excited Higgs bosons
due to their raditively induced couplings. 

In Sec. III we have considered only $2\rightarrow 4$ fermion processes in illustrating 
Higgs search via the Bjorken mechanism. The actual experimental search strategy \cite{LEP} 
is to look for $f_{2}=\nu$ and $f_1=b$ in eq. (37), which kills down the possibility of 
having a pseudoscalar boson coupling to $\bar{f}_{2}f_{2}$ pair. Such an event selection
mechanism necessarly establishes the existence of a scalar or set of scalars as the 
collider energy increases. However, when the collider energy gets higher and higher,
pairs of lightest neutralinos and charginos will be produced in association with the Higgs.
In this case $Z$ and $P$ couplings to  $\bar{f}_{2}f_{2}$ pair compete for $h_{s}\sim G$
as has been assumed throughout the work. In spite of all such supersymmetric particle 
production modes occuring in association with a CP--even  Higgs particle, the main 
search mecanishm for the CP--even Higgs remains to be 
the Bjorken process with $f_{2}=\nu$ and $f_1=b$ \cite{LEP}.

As a final point, it is obvious that $I(p^{2})$ described in the figures is not a
directly measurable quantity. However, it is highly useful in 
distinguishing the two models in terms of the di-fermion invariant mass.
Indeed, $I(p^{2})$ could be a useful tool in MONTE CARLO simulations of
these models in the next linear collider \cite{NLC}. Both the 
simulation studies (with more sophistcated numerical techniques) 
and analysis of the experimental data can be guided by the results 
of this analysis.

\section{Acknowledgements}
The author is grateful to Goran Senjanovic for highly stimulating
discussions and helpful comments. He also acknowledges the 
discussions with Alejandra Melfo.

\newpage
\begin{table}[htbp]
\begin{center}
\begin{tabular}{||c|c|c||}
Quantity&$Z^{\prime}$ Models&NMSSM\\ \hline 
$\lambda_{1}$&$G^{2}/8+{g_{Y'}}^{2}{Q}_{1}^{2}/2$& 
$G^{2}/8$\\ \hline
$\lambda_{2}$&$G^{2}/8+{g_{Y'}}^{2}{Q}_{2}^{2}/2$&
$G^{2}/8$\\ \hline
$\lambda_{S}$&${g_{Y'}}^{2}{Q}_{S}^{2}/2$&$k_{s}^{2}$\\ \hline 
$\lambda_{12}$&$-G^{2}/4+{g_{Y'}}^{2}{Q}_{1}{Q}_{2}+h_{s}^{2}$&
$-G^{2}/4+h_{s}^{2}$\\ \hline
$\lambda_{1S}$&${g_{Y'}}^{2}{Q}_{1}{Q}_{S}+{h_{s}}^{2}$&$h_{s}^{2}$\\ 
\hline
$\lambda_{2S}$&${g_{Y'}}^{2}{Q}_{2}{Q}_{S}+{h_{s}}^{2}$&$h_{s}^{2}$\\ 
\hline
$\lambda_{S12}$&0&$h_{s}k_{s}$\\
\hline
$\tilde{\lambda}_{12}$&$-h_{s}^{2}+g_{2}^{2}/2$&$-h_{s}^{2}+g_{2}^{2}/2$
\end{tabular}
\end{center}
\caption{\label{table:back1}
Explicit expressions for quartic couplings in $Z^{\prime}$ models and NMSSM.}
\end{table}
\begin{table}[htbp]
\begin{center}
\begin{tabular}{||c|c|c||}
$Z_{i}$&$g_{V}^{(i)}$&$g_{A}^{(i)}$\\ \hline
$Z_{1}$&$(\epsilon^{f}_{L}+\epsilon^{f}_{R})\cos\theta- 
\kappa({\epsilon^{\prime}}^{f}_{L}+{\epsilon^{\prime}}^{f}_{R})\sin\theta$& 
$(\epsilon^{f}_{L}-\epsilon^{f}_{R})\cos\theta-
\kappa({\epsilon^{\prime}}^{f}_{L}-{\epsilon^{\prime}}^{f}_{R})\sin\theta 
$\\ \hline 
$Z_{2}$&$(\epsilon^{f}_{L}+\epsilon^{f}_{R})\sin\theta+
\kappa({\epsilon^{\prime}}^{f}_{L}+{\epsilon^{\prime}}^{f}_{R})\cos\theta$&
$(\epsilon^{f}_{L}-\epsilon^{f}_{R})\sin\theta+
\kappa({\epsilon^{\prime}}^{f}_{L}-{\epsilon^{\prime}}^{f}_{R})\cos\theta$
\end{tabular}
\end{center}
\caption{\label{table:back2}
$Z_{i}f\bar{f}$ couplings in $Z^{\prime}$ models. Here $\kappa 
=g_{Y'}/G$. To obtain these couplings in NMSSM, one sets $g_{V,A}^{(2)}=0$, 
$\kappa=0$, and $\theta=0$.}
\end{table}
\begin{table}[htbp]
\begin{center}
\begin{tabular}{||c|c|c|c||}
$h_{k}$&$Z_{1}\,Z_{1}$&$Z_2\,Z_2$&$Z_1\,Z_2$\\\hline
$h_{1}$&$\frac{1}{\sqrt{6}}G\,M_{Z}$&$\frac{1}{\sqrt{6}}3\,G\,M_{Z} 
\rho^{2}$&0\\\hline 
$h_{2}$&0&0&$-\frac{1}{2}G\,M_{Z}\rho$\\\hline
$h_{3}$&$-\frac{1}{\sqrt{12}}G\,M_{Z}$&$\frac{1}{\sqrt{12}}3\,G\,M_{Z} 
\rho^{2}$&0 \end{tabular}
\end{center}
\caption{\label{table:back3}
$Z_{i}Z_{j}h_{k}$ couplings in $Z^{\prime}$ models. Here $\rho
=2(g_{Y'}/G)Q_{1}$.} 
\end{table}
\begin{table}[htbp]
\begin{center}
\begin{tabular}{||c|c|c||}
$h_{k}$&$Z\,Z$ (\mbox{NMSSM1}) &$Z\,Z$ (\mbox{NMSSM2})\\\hline
$h_{1}$&0&$\frac{1}{2}G\,M_{Z}$\\\hline
$h_{2}$&$\frac{1}{2}G\,M_{Z}$&0\\\hline
$h_{3}$&0&0\end{tabular}\end{center}
\caption{\label{table:back4}
$Z\,Z\,h_{k}$ couplings in NMSSM1 and NMSSM2 (see Eq. (19)).}
\end{table}
\begin{table}[htbp]
\begin{center}
\begin{tabular}{||c|c|c|c||}
$m_{h_{1}}$ (GeV)&$Z^{\prime}$ Models& \mbox{NMSSM1} & \mbox{NMSSM2}\\\hline 
Tree Level&121.8&126.0&121.8\\\hline
One-Loop &127.5&132.9&121.3\end{tabular}\end{center}
\caption{\label{table:back5}
Lightest Higgs mass in the models under concern.}
\end{table}
\newpage
\begin{center}
{\bf Figure Captions}
\end{center}
Fig. 1. Normalized cross section differential $I(x)$ for $Z^{\prime}$
models at tree- (dotted curve) and one-loop (full curve) levels.\\ 
Fig. 2. Same as in Fig. 1, but for NMSSM1.\\
Fig. 3. Same as in Fig. 1, but for NMSSM2.
\newpage 
\begin{figure}
\vspace{10cm}
\end{figure}
\begin{figure}
\vspace{12.0cm}
    \includegraphics{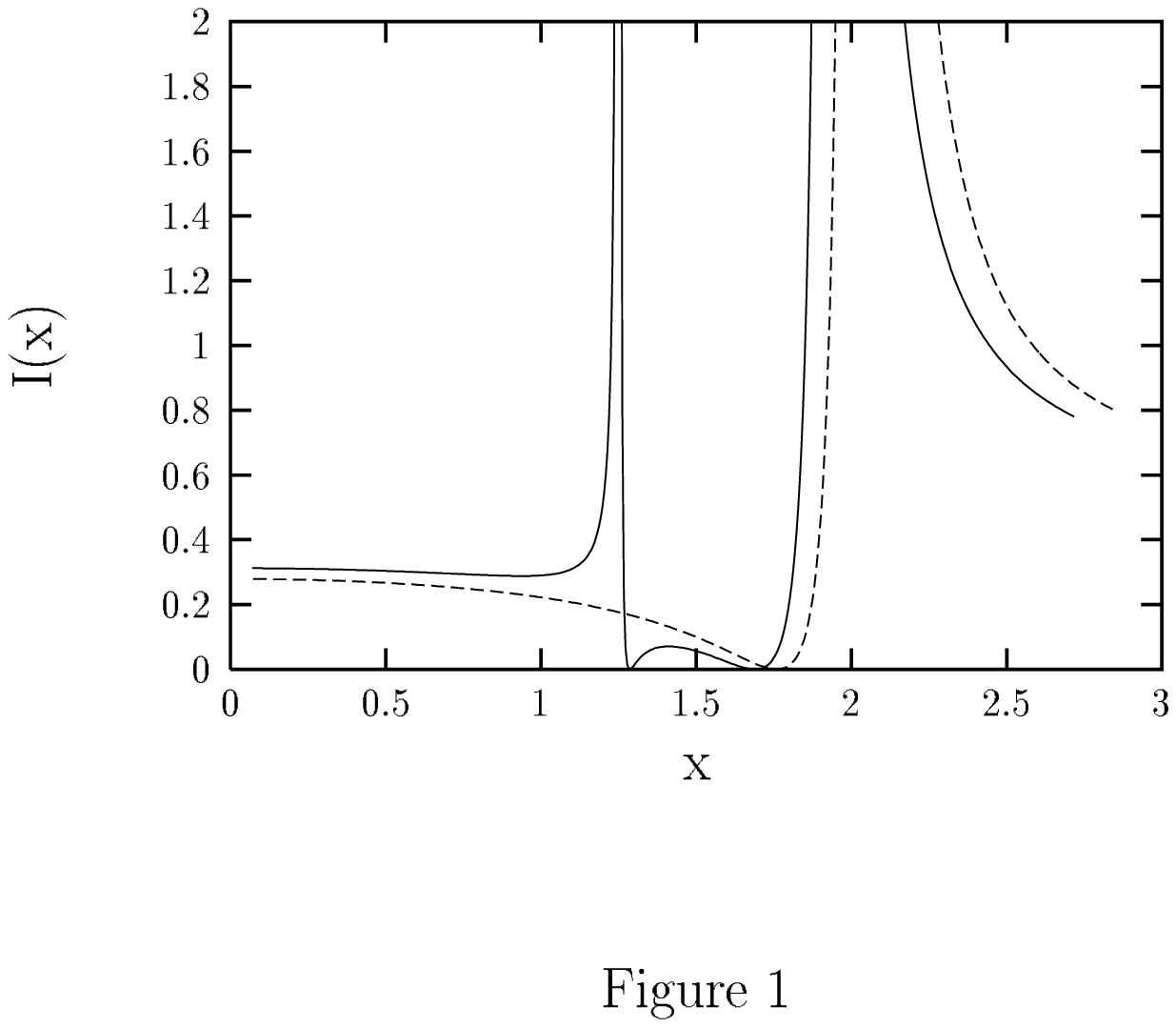}
    \vspace{-8.0cm}
\vspace{0.0cm}
\end{figure}
\begin{figure}
\vspace{10.0cm}
\end{figure}
\begin{figure}
\vspace{12.0cm}
    \includegraphics{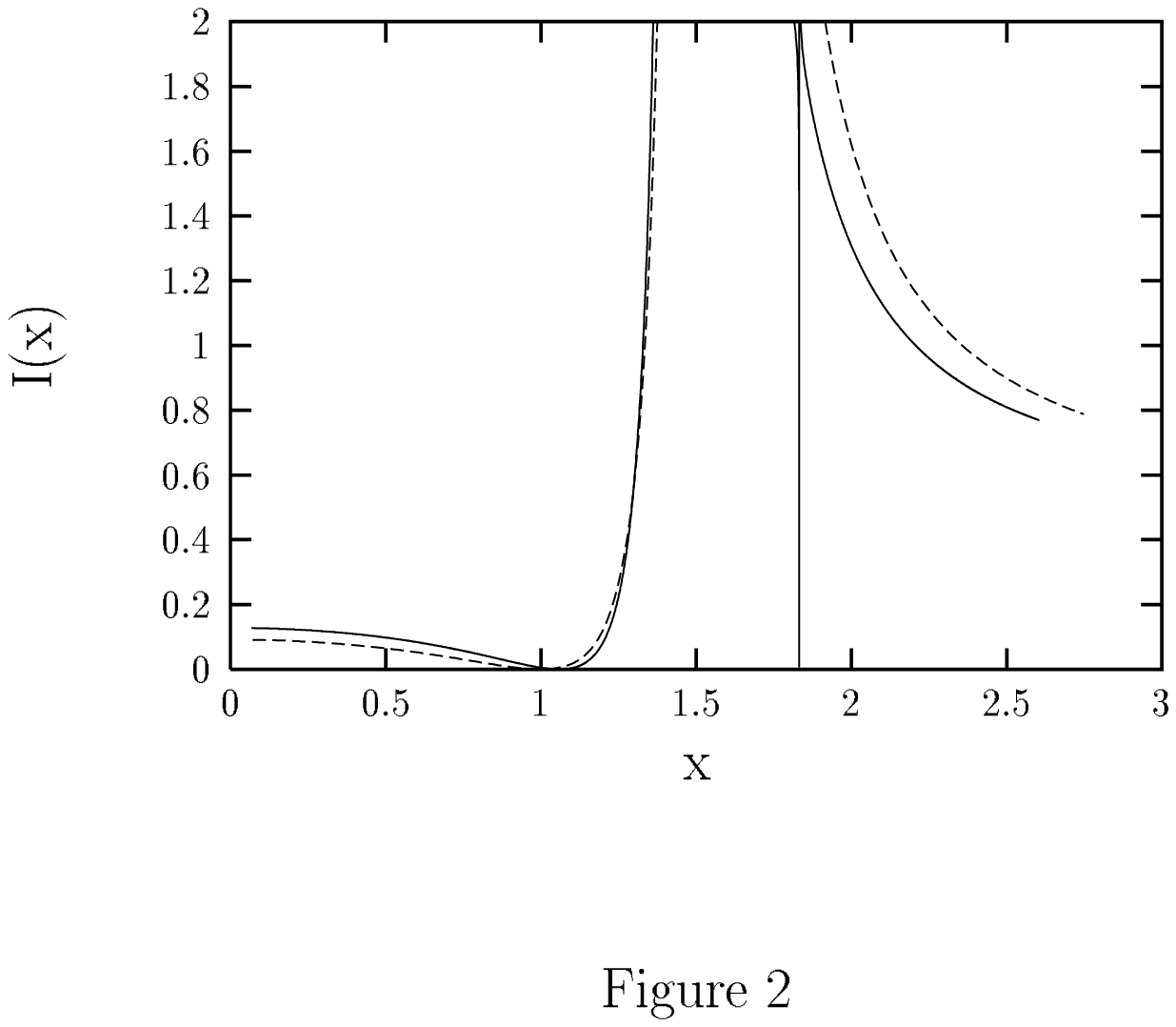}
    \vspace{-8.0cm}
\vspace{0.0cm}
\end{figure}
\begin{figure}
\vspace{10.0cm}
\end{figure}
\begin{figure}
\vspace{12.0cm}
    \includegraphics{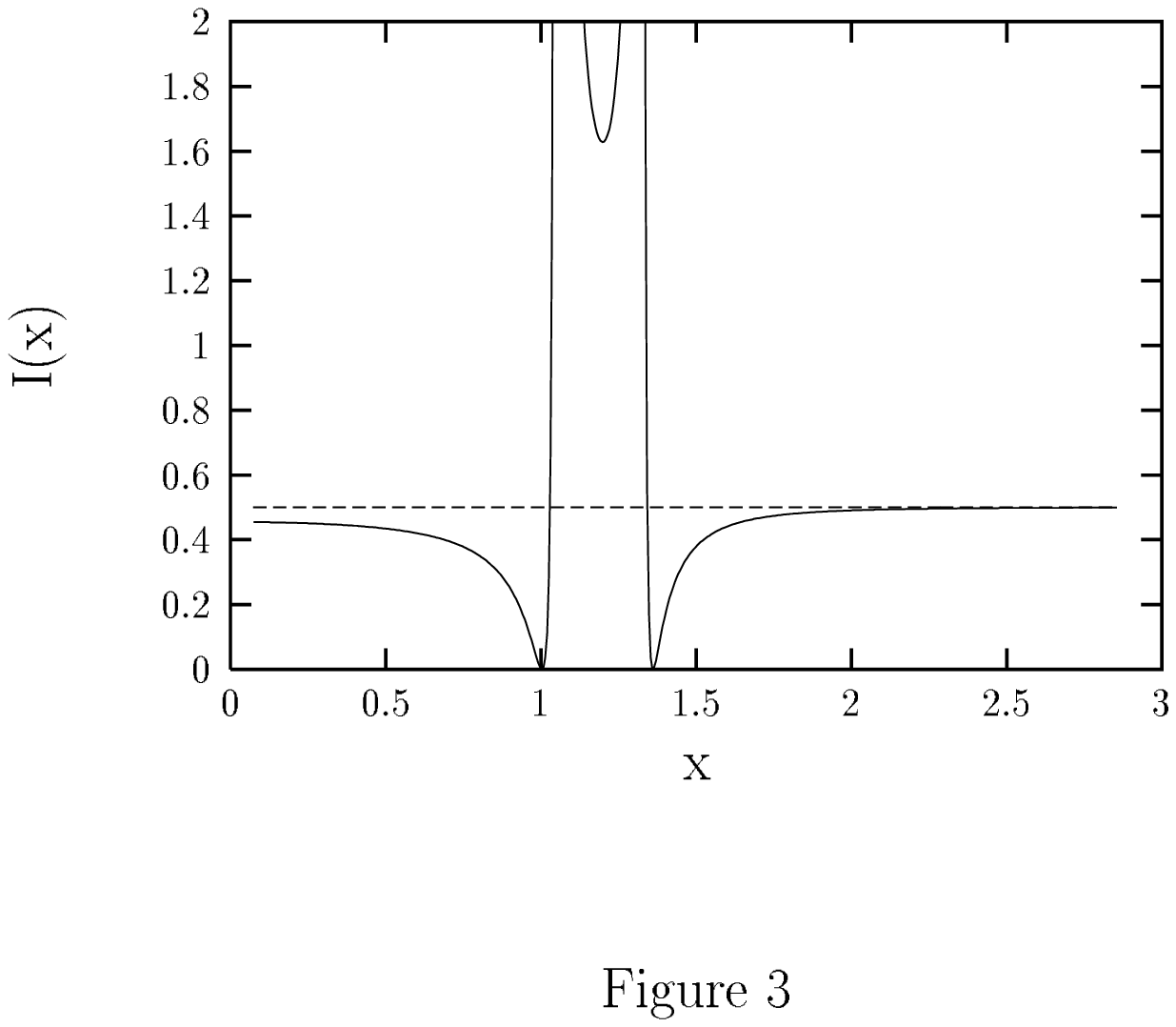}
    \vspace{-8.0cm}
\vspace{0.0cm}
\end{figure}
\end{document}